\documentclass[usenatbib]{mnras}

\usepackage{float}
\usepackage[T1]{fontenc}

\DeclareRobustCommand{\VAN}[3]{#2}
\let\VANthebibliography\thebibliography
\def\thebibliography{\DeclareRobustCommand{\VAN}[3]{##3}\VANthebibliography}

\usepackage{graphicx}	% Including figure files
\usepackage{amsmath}	% Advanced maths commands
\usepackage{amssymb}    % Extra maths symbols

\usepackage{newtxtext,newtxmath}
\title[Constraining the EoR With Highly Dispersed FRBs]{Constraining the Epoch of Reionization With Highly Dispersed Fast Radio Bursts}

\author[M. Pagano and H. Fronenberg]{
Michael Pagano\thanks{E-mail: michael.pagano@mail.mcgill.ca} and 
Hannah Fronenberg\thanks{E-mail: hannah.fronenberg@mail.mcgill.ca}
\\
Department of Physics and McGill Space Institute, McGill University, Montreal, QC H3A 2T8, Canada
}

\date{Accepted XXX. Received YYY; in original form ZZZ}

\pubyear{2021}

\begin{document}
\label{firstpage}
\pagerange{\pageref{firstpage}--\pageref{lastpage}}                                 
\maketitle
%We explore the effectiveness of using measurements of $\overline{\rm{DM}}$ bar to place constraints on the astrophysics and morphology of the EoR. We consider two observational scenarios 
%We show that the mean dispersion measure (DM) $\overline{\rm{DM}}$ of high redshift FRBs are sensitive to the astrophysics and morphology of the EoR.  
%We use a commonly used parametrization of EoR scenarios and study the dependence of DM on them.
\begin{abstract}
The period in which hydrogen in the intergalactic medium (IGM) is ionized, known as the Epoch of Reionization (EoR) is still poorly understood. The timing and duration of the EoR is expected to be governed by the underlying astrophysics. Furthermore, most models of reionization predict a correlation between the density and ionization field. Here we consider using the mean dispersion measure (DM) of high redshift Fast Radio Bursts (FRBs) as a probe of the underlying astrophysics and morphology of the EoR. To do this, we forecast observational scenarios by building mock data sets of non-repeating FRBs between redshifts $8\leq z \leq 10$. It is assumed that all FRBs have accompanying spectroscopic redshift measurements. We find that samples of 100 high redshift FRBs, in the above mentioned narrow redshift range, can rule out uncorrelated reionization at $68$\% credibility, while larger samples, $\geq 10^4$ FRBs, can rule out uncorrelated reionization at $95\%$  credibility. We also find 100 high 
redshift FRBs can rule out scenarios where the Universe is entirely neutral at $z = 10$ with $68\%$ credibility. Further with $\geq 10^5$ FRBs, we can constrain the duration $\Delta z$ of reionization (duration between mean ionized fraction 0.25 to 0.75) to $\Delta z = 2.0^{+0.5}_{-0.4}$, and the midpoint of reionization to $z = 7.8^{+0.4}_{-0.2}$ at $95$\% credibility. 

%that using Using Samples of 100 high redshift FRBs can also place broad constraints on the timing and duration of reionization at the $68\%$ CR level. We also show that large samples of FRBs ($\geq10^4$) in the above redshift ranges can further rule out uncorrelated reionization at $95$\% CR while measurements of $\geq10^5$ FRBs can place precise constraints on the timing and duration of reionization at the $95\%$ CR level.

%Jordan:I think it's great if you can close an abstract with a big punchy line -- right now this sentence is a bit of a downer, but if you just framed it in the affirmative instead it could be more powerful, e.g., "incorporating the full non-Gaussian errors in the likelihood could enable comparable inference with even fewer detections" or something like that.

%amount of FRBs measured in redshift bins $8 \le z \le 10$ can place bla constraints Using the We forecast the types of constraints that can be placed on the astrophysics and morphology of the EoR using measurements on $\overline{\rm{DM}}$. We find that measurement of *** high redshift FRBs during the EoR can constrain the morphology of reionization to $99\%$CR.
%We find that the morphology of the EoR affected the number density of free electrons in ionized regions and the astrophysics of the EoR 
\end{abstract}


%%%%%%%%%%%%%%%%%%%%%%%%%%%%%%%%%%%%%%%%%%%%%%%%%%

%%%%%%%%%%%%%%%%% BODY OF PAPER %%%%%%%%%%%%%%%%%%

\section{Introduction}

The Epoch of Reionization (EoR) is a transitional period in our Universe's history when the  neutral hydrogen (HI) making up the intergalactic medium (IGM) was ionized by the first generation of stars and galaxies. The Cosmic Microwave Background (CMB) has given us a peek into the early universe and measurements of quasars at $z < 7$ teach us about early galaxy evolution. Cosmic dawn and the EoR remain the missing piece of our understanding at $z_{\rm{CMB}} > z > 7$. Understanding this period not only provides insight into the very early universe, but also teaches us about the first generation of stars and galaxies. Despite its importance to our understanding of the Universe, the timing, mechanisms, and morphology of the EoR are poorly constrained. A number of observational probes have began making measurements of the EoR through the 21cm line \citep{GMRT,HERA,MWA,PAPER,EDGES}. The advantage of using this line as a direct probe IGM during the EoR is that neutral hydrogen is abundant in the early Universe and that by measuring the redshifting of this photon, we can trace primordial hydrogen along the line of sight. For a comprehensive review of 21cm cosmology, the reader is encouraged to read \cite{Morales&Wyithe}, \cite{FurlanettoOh&Briggs}, \cite{Pritchard&Loeb}, \cite{LoebFurlanetto2013} and \cite{AdawgsReview}. 

21cm cosmology, however, does not come without its challenges. Making a detection of the 21 cm line during the EoR is exceptionally difficult since the frequency of the line is redshifted into the 50-300 MHz range \citep{AdawgsReview}. Systematics, radio frequency interference (RFI), galactic synchrotron emission, and radio bright sources have made the 21 cm signal difficult to measure, and thus limit our ability to constrain the astrophysics during this epoch \citep{AdawgForegrounds}. As a result, many look to other probes of the EoR.

%The Epoch of Reionization (EoR) is a transitional period in our Universe's history when the  neutral hydrogen (HI) making up the intergalactic medium (IGM) was ionized by the first generation of stars and galaxies. This period is the missing piece of our understanding of the Universe between the Cosmic Microwave Background (CMB) and measurements of high redshift quasars at $z < 7$. It thus has important consequences for our understanding of galaxy formation. Despite its importance to our understanding of the Universe, the timing, mechanism and morphology of the EoR are poorly constrained. A number of observational probes have began making measurements of the EoR through the hyperfine transition of the electron in hydrogen; the 21cm line\hannahsf{Add ref of EDGES,HERA,MWA,PAPER}. The advantage of using this line as a probe of the EoR is that primordial neutral hydrogen is abundant in the early Universe and that by measuring the redshifting of this photon, we can trace the primordial hydrogen along the line of sight. The photon is part of the radio spectrum and is measured in contrast to the Cosmic Microwave Background (CMB). For a comprehensive review of 21cm cosmology, the reader is encouraged to read \cite{AdawgsReview}. Systematics, radio frequency interference (RFI), and radio bright foregrounds have made the 21cm signal difficult to measure and thus limited our ability to constrain the astrophysics during this Epoch. As a result, other probes of the EoR have been proposed.

Fast Radio Bursts (FRBs) are a class of bright, millisecond duration, radio transients that have been detected at frequencies ranging from 110 MHz to 1.5 GHz and whose dispersion measures (DMs) lie between 110 and 2600 pc cm$^{-3}$ \citep{Lofar,Petroff2016}. Thanks to current and upcoming broad-band wide-field-of-view instruments, such as the Canadian hydrogen Intensity Mapping Experiment (CHIME; \cite{CHIMEtelescope}), the hydrogen Intensity and Real-time Analysis eXperiment (HIRAX; \cite{HIRAX}), Five-hundred metre Aperture Spherical Telescope (FAST; \cite{FAST}), Australian Square Kilometer Array Pathfinder (ASKAP; \cite{ASKAP}) and the Square Kilometer Array (SKA; \cite{SKA}), we have seen a large increase in the number of FRBs detected. It is estimated that when SKA is online, its event detection rate may be as high as 
$\sim$1000 FRBs sky$^{-1}$ day$^{-1}$ \citep{SKArate}.

Since their discovery by \cite{Lorimer2007}, on the order of $10^3$ FBRs have been observed, over 1000 of which by CHIME alone. While one source has now been localized within the Milky Way (\cite{CHIMEmagnetar}), the vast majority of FRBs remain extragalactic sources and can thus probe out to cosmological distances \citep{Dolag,Katz}. Many question remain about the astrophysical origin of these bursts as well as their intrinsic distribution out to high DM \citep{Wenbin2020,Wenbin,FRBtheoryCat}. 

While the progenitor of FRBs remains unknown, the DMs of these bursts are, on the contrary, well understood and could thus prove yet another direct probe of the IGM during the EoR. The DM of an FRB is defined as the integrated column density of free electrons along the line of sight from source to observer. As the FRB travels through the IGM, it experiences a frequency dependent time delay, $\Delta t \propto \nu^{-2}\textrm{DM}$. $\textrm{DM}$ is given by
\begin{equation}
\label{eq:DM_intro}
\textrm{DM}(\mathbf{x},z) = \int \frac{n_e(\mathbf{x},z)}{1+z} dl ,
\end{equation}
and where $dl$ is the line element along the light of sight, $n_e(\mathbf{x},z)$ is the free electron density at comoving position $\mathbf{x}$ and redshift $z$. Measuring the DM of an FRB at redshift $z$ can therefore probe the integrated number density of free electrons along the line of sight in the IGM. To evaluate \ref{eq:DM_intro} for each reionization scenario, we express the line element $dl$ in terms of the Hubble parameter
\begin{equation}
    dl = c dt = \frac{-c dz}{H(z)(1 + z)}
\end{equation}
where $H(z)$ is given in terms of the $\Lambda$CDM parameters through
\begin{equation}
    H(z) = H_0\sqrt{\Omega_m ( 1 + z )^3 + \Omega_\Lambda} \equiv H_0 E(z).
\end{equation}
The free electron number density in the IGM can be written as a function of the ionization and density field, 
\begin{equation}
\label{eq:ne_intro}
n_e = \frac{\textrm{f}_{\rm{H}} \textrm{f}_{\rm{IGM}} \Omega_m \rho_{0}(z)}{m_{\rm{H}}} (1 + z)^3 ( 1 + \delta(\textbf{x},z) )x_{\rm{HII}}(\textbf{x},z), 
\end{equation}
where $\textrm{f}_{\rm{H}}$ is the fraction of baryonic matter that is hydrogen, $\textrm{f}_{\rm{IGM}}$ is the fraction of hydrogen that is found in the IGM, m$_{\rm{H}}$ is the mass of hydrogen and $\rho_0$ is the mean density of the IGM at redshift $z$. 
%The electron number density $n_e$ in the IGM can be written as 
%\begin{equation}
%\label{eq:ne_intro}
%    n_e = f_{\rm{H}} \Omega_b \rho_{0}(z)[ 1 + \delta(\mathbf{x},z) ]    x_{\rm{HII}}(\mathbf{x},z)
%\end{equation}
%\rho_0(\mathbf{x},z)
%where $\delta(\mathbf{x},z)$ is the baryonic overdensity, %$x_{\rm{HII}}(\mathbf{x},z)$ is the ionization fraction of hydrogen, and $f_H$ the hydrogen fraction of baryons and $\rho_0$ is the mean density at redshift $z$. 
The dispersion measure of FRBs detected after the EoR can be approximated to be $x_{\rm{HII}}(\mathbf{x}) = 1$, i.e. the IGM is entirely ionized. Note that helium reionization does increase the number density of free electrons at low redshift ($z \sim 2 $), however this is independent of the reionization model. Therefore, we do not take this into account in our models for Equation \ref{eq:DM_intro}. For interested readers, there is a growing body of literature on constraining helium reionization using the DMs of FRBs \citep{Zheng_2014,Caleb_2019,Linder_2020,bhattacharya2020fast}. Unlike at low redshift, the ionization field $x_{\rm{HII}}$ during the Epoch of Reionization is patchy, composed of regions of ionized bubbles and neutral regions, whose placements and evolution depend highly on the astrophysics governing reionization. In this case, the number density of electrons $n_e$ will dependent on the state of the ionization field. High redshift FRBs detected during the Epoch of Reionization will  therefore be sensitive to the astrophysics that have imprinted itself onto  $x_{\rm{HII}}$. 

Furthermore, referring to Equation \ref{eq:ne_intro}, the number density of free electrons is dependent on the product of the density and ionization field $ x_{\rm{HII}}\delta$. The method in which the ionization field $x_{\rm{HII}}$ maps to the underlying density field $\delta$ is known as the density-ionization correlation, which affects the morphology of the EoR. The cross term in Equation \ref{eq:ne_intro} contains information of the density-ionisation correlation, which affects the observed DM of an FRB. Most EoR models predict morphologies where the ionized regions are not random with respect to the underlying density field. Instead, there are two extreme ways in which the ionization field couples to the density field. The density field can be positively correlated to the ionization field. In this scenario, overdense regions correspond to high ionization fraction. In this model, ionizing sources ionize their immediate surroundings before ionizing the lower density regions of the IGM. We say that reionization happens `inside-out'. The second extreme model is the scenario where the underlying density field $\delta$ is negatively correlated with the ionization field $x_{\rm{HII}}$. In this scenario, the ionizing sources first ionize the low density regions before ionizing the high density regions. We say that reionization happens `outside-in'. In this model, the high density regions in $\delta$ correspond to regions of low ionization fraction in $x_{\rm{HII}}$. This model usually requires the recombination of hydrogen atoms in high density regions to dominate the effects of UV ionization \citep{Choudhury, Miralda2000, WatPrit}. Outside-in morphologies can also be achieved by having reionization driven by x-ray photons, which can more easily ``leak" into the underdense regions of the IGM \citep{AndreiXray, xraymodelsMirabel}. It is also possible for reionization to unfold as a combination of both inside-out and outside-in, in which case, the correlations between $\delta$ and $x_{\rm{HII}}$ are statistical combination of the inside-out and outside-in models \citep{FurlanettoOhCombination2005, MadauHaardt2015}. 

The way in which the mean DM depends on the broad timeline of the reionization history has been previously studied \citep{Hashimoto2021,Zhang+2021, Beniamini_2021,Macquart2018}. Most recently, \cite{Hashimoto2021} show that one year's worth of observing with SKA phase 2 can reveal our cosmic reionization history, and \cite{Beniamini_2021} show that both DM and the differential FRB source count distribution prove useful probes of reionization even with limited redshift information. In this paper we build on the techniques outlined by these authors by performing a study of how the morphology, astrophysics and evolution of the EoR affect the mean DM of high redshift FRBs. We use a set of astrophysical and morphological parameters to bracket the physical range of EoR scenarios and study how DM--z probability distributions and the mean DM of FRBs at each redshift depend on these parameters. We then forecast the types of constraints that we can place on the EoR using measurements of the DMs of high redshift FRBs. Since FRBs at high redshift have yet to be observed, we create a mock sample of highly dispersed FRBs under a fiducial reionization model and forecast the type of constraints one can place on the astrophysics and morphology of the EoR given such a measurement. We perform this forecast with $10^2$, $10^4$, and $10^5$ high DM samples.

This paper is structured as follows. In Section \ref{sec:SimulationParams} we describe the astrophysical and morphological parameters used in our simulation to bracket the physical range of EoR scenarios. In Section \ref{sec:models} we discuss how the mean DM and the DM probability distributions of high redshift FRBs depend on these parameters. In Section \ref{sec:forecasts}, we describe our fiducial reionization model, the mock FRB measurements made for this reionization scenario. We forecast the constraints that can be placed on the EoR parameters using such a measurement and in Section \ref{sec:Results} we present the results. We summarize our conclusions in Section \ref{sec:Conclusion}. Throughout this work we set the $\Lambda$CDM parameters to $\sigma_8 = 0.81$, $\Omega_m = 0.31$, $\Omega_b = 0.048$, $h = 0.68$ \citep{Planck}.

\begin{table*}
\caption{Summary of Reionization Parameters and FRB Observables \label{tab:params_summary}}
\begin{center}
\begin{tabular}{|c|c|c|} 
\hline
 Symbol & Parameter Name & Description/Definition \\ 
\hline\hline
$z_{\rm{EoR}}$ & Reionization Redshift & The redshift indicating the onset of reionization.\\
\hline
  $\beta$ & Morphological Parameter &  Determines the correlation between $\delta$ and $x_{\rm{HII}}$\\
\hline
 $M_{\rm{turn}}$  & The turnover mass &  Halo mass scale in which star formation is efficient \\
  \hline
 $\zeta$ & Ionizing Efficiency & Number of ionizing photons released per stellar baryon\\ 
\hline
  $R_{\rm{mfp}}$ & Radius of The Mean Free Path & Maximize size of the ionized regions\\ 
  \hline
$\rm{DM}(\textbf{x},z)$ & Dispersion Measure & DM of an individual FRB at redshift z along a single line of sight.\\ 
\hline
$\overline{\rm{DM}}(z)$ & Mean Dispersion Measure & Mean DM of a collection of FRBs observed at redshift $z$\\ 
\hline
\end{tabular}
\end{center}
\end{table*}

\section{Simulation}
\label{sec:SimulationParams}
To generate density and ionization boxes representative of different EoR models we use \texttt{21cmFAST} package \citep{21cmFAST}. Density fields are obtained through the Zeldovich approximation while ionization and halo boxes implement the excursion set formalism of \cite{FZH}. For further details about how \texttt{21cmFAST} generates reionization models see cite{21cmFAST}. Throughout this paper we use high resolution boxes of  $800^3$ voxels corresponding to a comoving side length of $300\,\textrm{Mpc}$ and coarser boxes of $200^3$ voxels corresponding to the same comoving side length. 

\subsection{EoR Parameters}
We use \texttt{21cmFAST} to generate different EoR scenarios by varying a number of adjustable parameters which encapsulate variations in the detailed astrophysics of reionization. We bracket the physical range of EoR scenarios by adjusting the parameters $M_{\rm turn}$, $R_{\textrm{mfp}}$, and $\zeta$. Physically, the turnover mass, $M_{\rm turn}$, determines the mass of a halo in which star formation is efficient. Values of $M_{\rm turn} \simeq 5\times 10^8 M_\odot$ correspond to a virial temperature of $T_{\textrm vir} \simeq 10^4$. Values below $M_{\rm turn}$ have exponential suppression in star formation. Roughly, this sets the mass scale for the ionizing sources. The unitless astrophysical parameter $\zeta$, determines the ionizing efficiency of the sources. This parameter is an amalgamation of other parameters which describe the small scale astrophysics of the UV sources. A large value of $\zeta$ will imply more ionizing photons per stellar baryon, while a smaller ionizing efficiency will entail less ionizing photons are emitted for each ionizing source. The cutoff-radius $R_{\textrm{mfp}}$ sets the maximum size of the ionized bubbles \citep{SobacchiAndrei}. Variation of these parameters effect the timing and duration of reionization, and have been studied in previous works \citep{AdrianMCMC, KernEtAl,LiuParsonsParams, EwallWiceForecast2016, Park}. For this work, we use these parameters to generate a wide variety of EoR models that bracket physical scenarios. These parameters operate under an inside-out reionization formalism in which the density field $\delta$ is correlated with the ionization field $x_{\rm {HII}}$ and therefore do not capture the different $\delta$ $x_{\rm {HII}}$ correlations indicative of different EoR morphologies. In the next Section we introduce a parametrization that extends the physical scenarios bracketed by the astrophysical parameters to EoR morphologies of arbitrary ionization-density correlations. 

\subsection{Morphological Parametrization of the EoR }
\label{sec:beta_parametrization}
\begin{figure}
    \includegraphics[width=0.49\textwidth]{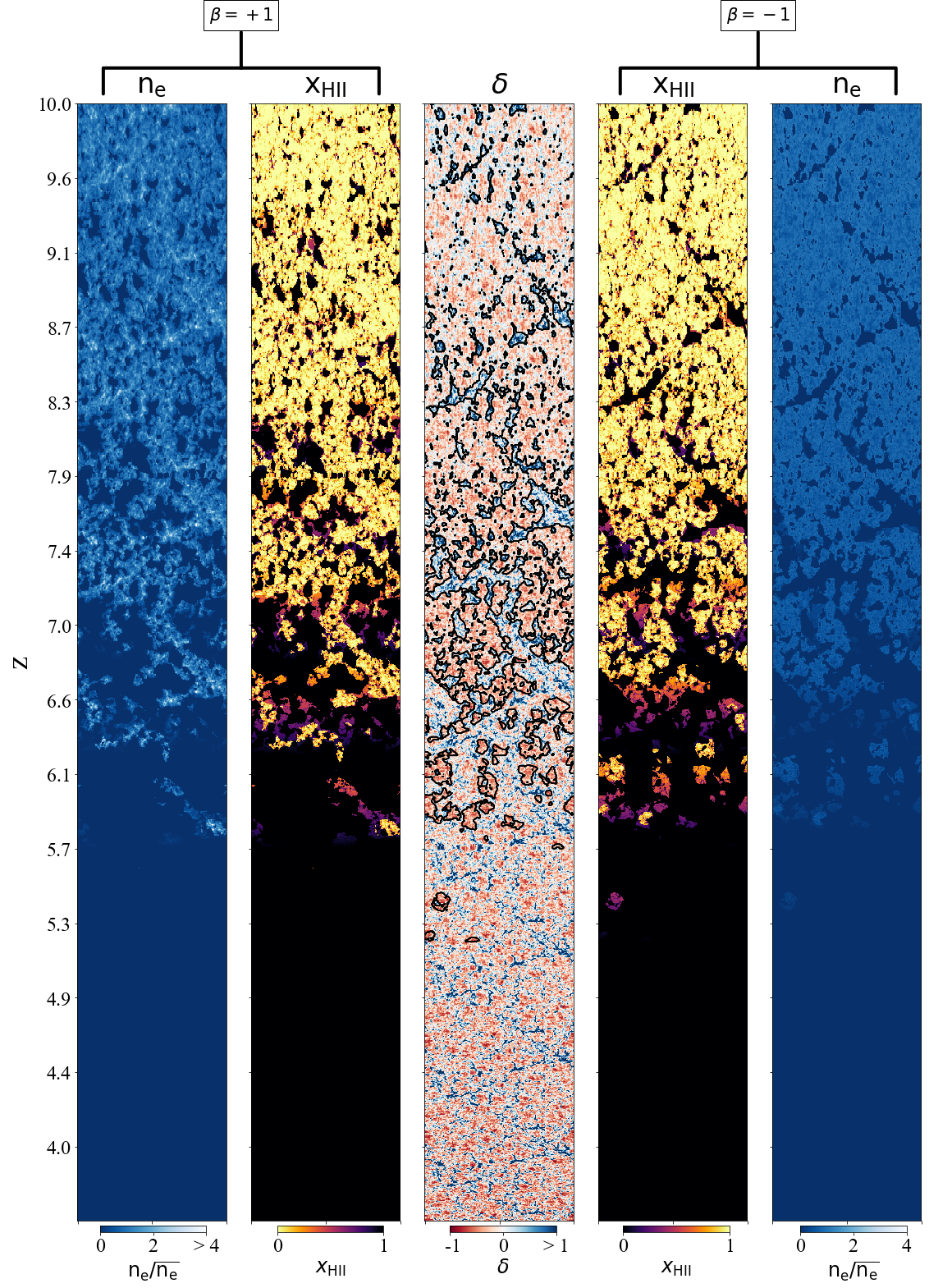}
    \caption{From the center of the Figure , outward: Lightcones of the density field $\delta$, ionization fraction $x_{\rm{HII}}$ and free electron field $n_e$ for the case of inside-out reionization (left three boxes) and outside-in reionization (right three boxes). Inside-out reionization (left) leads to a higher free electron number density $n_e$ in the ionized bubbles since the density field (center) couples to the ionized regions in $x_{\rm{HII}}$ compared to outside-in models (right) where the underdense regions in $\delta$ couple to the ionized regions in $x_{\rm{HII}}$.}
\label{fig:lightcones}
\end{figure}

 To simulate EoR scenarios where the density field and ionization field are correlated by some arbitrary amount, we use the $\beta$ parametrization introduced in \cite{me!}. This parameter continuously tracks the correlation between $x_{\rm {HII}} \delta$. We briefly describe this parametrization here. The $\beta$ parameter has bounds $-1 \le \beta  \le 1 $ and controls the amount of correlation between $x_{\rm HII}$ and $\delta$. The sign of $\beta$ indicates the overall sign of the correlation between  $x_{\rm HII}$ and $\delta$. Positive values of $\beta$, indicate a positive correlation between density and ionization fields, and so overdense regions in $\delta$ couple to regions of high ionization fraction in $x_{\rm HII}$. This sign of correlation is indicative of inside-out reionization scenarios, where the overdense regions of the IGM are first to be ionized. Conversely, negative signs of $\beta$ indicate an overall negative correlation between ionization field and density field so that overdense regions in $\delta$ correspond to regions of low ionized fraction of hydrogen. This is indicative of outside-in reionization, where overdense regions of the IGM are last to be ionized. The magnitude, $|\beta|$, indicates how strong that correlation sign is between ionization and density fields. A value of $\beta = 0$ indicates a random placement of the ionized regions, in which case there is no correlation between ionization and density fields. As we increase $\beta$ from $0$ to $1$, the relative likelihood of finding overdense regions of $\delta$ corresponding to ionized regions in $x_{\rm HII}$ increases, until finally at $\beta$ of $1$, all overdense regions in $\delta$ always correspond to regions of ionized hydrogen. Similarly, as we decrease $\beta$ from $0$ to $-1$, the relative likelihood of finding overdense regions of $\delta$ corresponding to ionized regions in $x_{\rm HII}$ increases until at $\beta = -1$, all overdense regions in $\delta$ correspond to regions of low $x_{\rm HII}$. The intermediate, non-extreme values of $\beta$, i.e. $-1 < \beta < 1$ indicate reionization scenarios that contain the statistics of both inside-out and outside-in. Figure \ref{fig:lightcones} demonstrates the affect of inside-out, or outside-in reionization, on the free electron number density $n_e$ in the IGM. For a more detailed discussion on this parametrization, the reader is encouraged to read \cite{me!}. Table \ref{tab:amps1} summarizes the terminology used to describe the type of correlation as well as the model to which it pertains.
 
 \begin{table}
\caption{Lexicon for physical models and their respective correlations \label{tab:amps1}}
\begin{center}
\begin{tabular}{|c|c|c|} 
\hline
$\beta$ & Moniker for Field correlations  & Physical Model \\ 
        &   $x_{\rm HII}$ $\delta $   &               \\
\hline\hline
1 &  Correlated &  Inside-out\\
\hline
  $1 < \beta < 0$ & Increasingly correlated & Mostly inside-out\\ 
\hline
 $0$  & Uncorrelated & Random\\ 
  \hline
 $0 < \beta < -1$ & Increasingly anti-correlated & Mostly outside-in\\ 
\hline
  $-1$ & Anti-correlated & Outside-in\\ 
  \hline
\end{tabular}
\end{center}
\end{table}

%An FRB located at redshift $z$ and position $\mathbf{x}$ has observed DM 
%\begin{equation}
%\label{eq:DM_total}
%\textrm{DM}_{\rm{obs}}(\mathbf{x},z) = \int \frac{n_e(\mathbf{x},z)}{1+z} dl ,
%\end{equation}
%where $dl$ is the comoving line element along the line of sight and $n_e(\mathbf{x}, z)$ is the free electron density. 
\subsection{Dispersion Measure}
\label{sec:DM}
The observed dispersion in Equation \ref{eq:DM_intro} is sensitive to all sources of free electrons encountered by the radio burst as the electromagnetic wave travels from source to observer. This includes the free electrons found within the host galaxy as the FRB leaves the source, as well as the free electrons encountered in the Milky Way (MW) as the FRB arrives to the observer. The FRB also has exposure to the free electrons found in the circumgalactic medium (CGM) and IGM. We split Equation \ref{eq:DM_intro} into its respective components 

%\hannahsf{Maybe we should call $DM_{TOT}$ $DM_{obs}$ since that seems to be what's common in the lit. we should also split up ISM into $DM_{MW}$ and $DM_{host}$. Then we can say we can remove the ISM components, (MW and host) and what's left over is DM cosmo which is CGM+ IGM. Then we say that CGM is small so really we approximate DMcosmo=DM IGM}
%\textrm{DM}_{\rm{ISM}}(\mathbf{x},z)

\begin{equation}
\label{eq:DM_components}
\textrm{DM}_{\rm{obs}}(\mathbf{x},z) = \textrm{DM}_{\rm{host}} + \textrm{DM}_{\rm{MW}} + \textrm{DM}_{\rm{CGM}}(\mathbf{x},z) +\textrm{DM}_{\rm{IGM}}(\mathbf{x},z) .
\end{equation}
The dispersion $\textrm{DM}_{\rm{IGM}}(\mathbf{x},z)$ is due to the free electrons found in the IGM between the FRB source and the observer. This is the DM attributed to cosmic reionization and is the DM of interest in order study the evolution of $x_{\rm{HII}}$. The DM attributed to the host galaxy, MW and CGM are subject to uncertainties surrounding the gas dynamics within these regimes, and as a result make them difficult to model. We treat them as contaminants in our measurement of the contributions of $\textrm{DM}_{\rm{IGM}}$ to $\textrm{DM}_{\rm{obs}}$. The DM contribution due to the interstellar medium (ISM)
of intervening galaxies have also been shown to be negligible \citep{ISMinterveningGalaxiesNeg}. The inhomogeneity of $x_{\rm{HII}}$, $\delta$ and the gas dynamics as a function of position $\mathbf{x}$ make it unreasonable to draw conclusions on the state of the IGM through a single line of sight. We instead compute the mean value of $\textrm{DM}_{\rm{obs}}$ due to all sightlines. This removes single line of sight fluctuations in $\delta$ and $x_{\rm {HII}}$ as well as averages over the contributions due to the CGM, 
\begin{equation}
\label{eq:DM_components_mean}
\overline{{\textrm{DM}}}_{\rm{obs}} = \overline{\textrm{DM}}_{\rm{host}} +\overline{\textrm{DM}}_{\rm{MW}} +  \overline{\textrm{DM}}_{\rm{CGM}} +\overline{\textrm{DM}}_{\rm{IGM}} .
\end{equation}
Studies such as \cite{KeatingPenDMOffset} model the DM contribution of the Milky Way, which is something which may be possible to accurately account for in the future. Other studies have found that the photon incurs an average DM of $\overline{\textrm{DM}}_{\rm{MW}} \sim 200$pc cm$^{-2}$ when leaving the MW and host galaxy \citep{DM_HOST, JonSeivers}. We treat the average contribution of the MW and host galaxy to $\overline{{\textrm{DM}}}_{\rm{obs}}$ as an offset $\sim 200$pc cm$^{-2}$. 
\begin{equation}
\label{eq:DM_remove_offset}
\overline{{\textrm{DM}}}_{\rm{obs}} - (\overline{\textrm{DM}}_{\rm{host}} +\overline{\textrm{DM}}_{\rm{MW}}) =  \overline{\textrm{DM}}_{\rm{CGM}} +\overline{\textrm{DM}}_{\rm{IGM}} .
\end{equation}
We assume high redshift FRBs have $\overline{{\textrm{DM}}}_{\rm{obs}}$ dominated by the IGM, we neglect the contribution due to the CGM. The remaining fluctuations in DM are attributed to cosmic reionization. Henceforth we refer to  $\textrm{DM}_{\rm{IGM}}$ as  $\textrm{DM}_{\rm{obs}}$. This model isn't meant to be overly realistic, we intend to capture first order effects due to $\textrm{DM}_{\rm{IGM}}$. In order for precise measurements to be made of the impact that the EoR has $\overline{\textrm{DM}}_{\rm{obs}}$, a method to subtract out the effects due to the CGM needs to be studied. We leave such a study to future work. 
The mean DM of a high redshift FRB observed at redshift $z$ due to free electrons in the IGM is then evaluated using Equation \ref{eq:DM_intro} as 
\begin{equation}
    \label{eq:DM_obs_z_mean_final}
    \overline{{\textrm{DM}}}_{\rm{obs}}(z) = -\int c dz \frac{ \textrm{f}_{\rm{H}} \textrm{f}_{\rm{IGM}} \Omega_m \rho_{\rm 0} (1 + z)}{m_{\rm{H}} H_0 E(z)}\left( \overline{x}_{\rm{HII}}(z) + \overline{\delta x}_{\rm{HII}}(z)\right) .
\end{equation}

The $\overline{{\textrm{DM}}}_{\rm{obs}}$ of high redshift FRBs will be proportional to the mean ionization fraction $\overline{x}_{\rm{HII}}$ of the IGM as well as to the mean product $\overline{x_{\rm{HII}}\delta}$. This cross term captures the density-ionization correlation of the EoR which describes how the underlying density field $\delta$ couples to the ionization field $x_{\rm{HII}}$. The $\beta$ parameter quantifies the different possibilities of this correlation. The astrophysics of the EoR, affect both of these terms. Since the astrophysics set the size and morphological features of the ionization field $x_{\rm{HII}}$, they too have consequences for $\overline{\rm{DM}}$ . In addition, the astrophysics of the EoR determine the onset and duration of the EoR, i.e. they determine the mean ionization fraction $\overline{x}_{\rm{HII}}$ at each redshift $z$. Previous studies have looked at how broad modeling the mean ionization fraction $\overline{x_{\rm{HII}} \delta}$ to redshift affects the observed DM of high redshift FRBs, i.e. the $\overline{x_{\rm{HII}}\delta}$ term in Equation \ref{eq:DM_obs_z_mean_final} \citep{Zhang+2021}. Here we build on that by including both terms and studying how the detailed astrophysics as well as density-ionization correlation affect both terms. For readers who are less familiar with the parameters that have been discussed, they are summarized in Table \ref{tab:params_summary} for easy reference. In the following Sections, we study how the astrophysical parameters, and the $\beta$ parameter, which parameterizes the density ionization correlation, affect $\overline{{\textrm{DM}}}_{\rm{obs}}$.   

In the following section, we evaluate Equation \ref{eq:DM_obs_z_mean_final} by simulating 10$^5$ sightlines, computing the individual DM of each sightline, and by averaging the DMs. This is done for each reionization model. 
\begin{figure*}
  \includegraphics[width=0.98\textwidth]{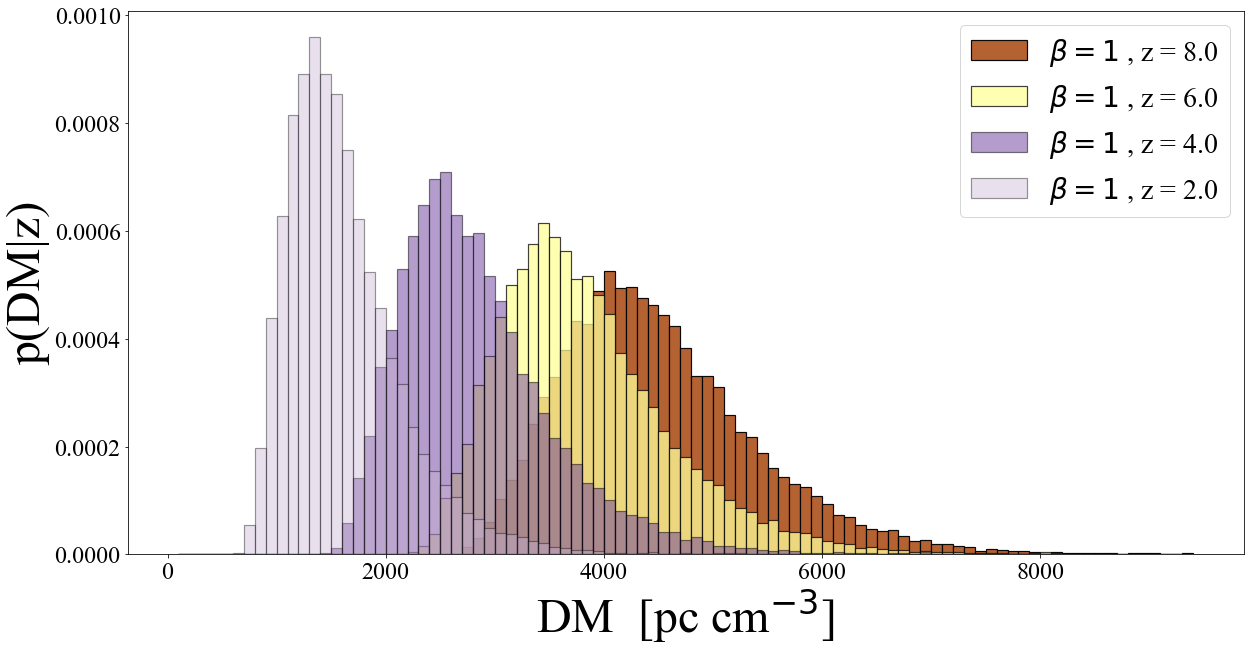}
  \caption{Redshift evolution of the DM probability distributions for our fiducial reionization scenario $\beta = 1$, $\zeta = 25$, $M_{\rm{turn}} = 5\times 10^8$M$_\odot$ and $R_{\rm{mfp}} = 30$Mpc. At higher redshift the relative probability of high DM sightlines increases.}
\label{fig:Prob_DM_single}
\end{figure*}

\section{Models}
\label{sec:models}

%From Section \ref{sec:DM} we neglect the fluctuations of $\overline{\rm{DM}}$ due to the ISM and CGM and consider only the $\overline{\rm{DM}}$ attributed to comsic reionization. We therefore treat fluctuations in $\overline{\rm{DM}}$ as fluctuations in $n_e$ produced by cosmic reionziation. 

Equation \ref{eq:DM_obs_z_mean_final}, states that $\overline{{\textrm{DM}}}_{\rm{IGM}}$ of an FRB depends on the cumulative of both the mean ionization fraction, $\overline{x}_{\rm{HII}}$, and the density-ionization correlation, $\overline{x_{\rm{HII}}\delta}$, along the line of sight. If $z_{\rm EoR}$ is the redshift in which the IGM becomes increasingly neutral, then $\overline{x}_{\rm{HII}} = 1 $ for all $z < z_{\rm EoR}$ and so the relationship between $\overline{{\textrm{DM}}}_{\rm{IGM}}$ and $z$ is linear up until the onset of reionization \citep{Liam}. The linear relationship breaks down at $z_{\rm EoR}$ since $\overline{x}_{\rm{HII}}$ decreases rapidly due to the increasingly neutral IGM. As a result, the EoR produces a flattening of $\overline{\rm{DM}}$ for high redshift FRBs. The shape and positioning of this flattening is highly dependent on the onset, duration and morphology of reionization. In this Section we use the astrophysical and correlation parameters to study how EoR models affect the DM of FRBs observed in the EoR. We consider the distribution of DM at each $z$ of the individual DM sightlines as a function of the astrophysics and morphology of the EoR. In Section \ref{sec:Results}, we forecast that the constraints that can be placed on these parameters through measurement of high redshift FRBs.

 \subsection{DM Distributions}
\label{subsec:DM_Distributions}
\begin{figure*}
  \includegraphics[width=0.99\textwidth]{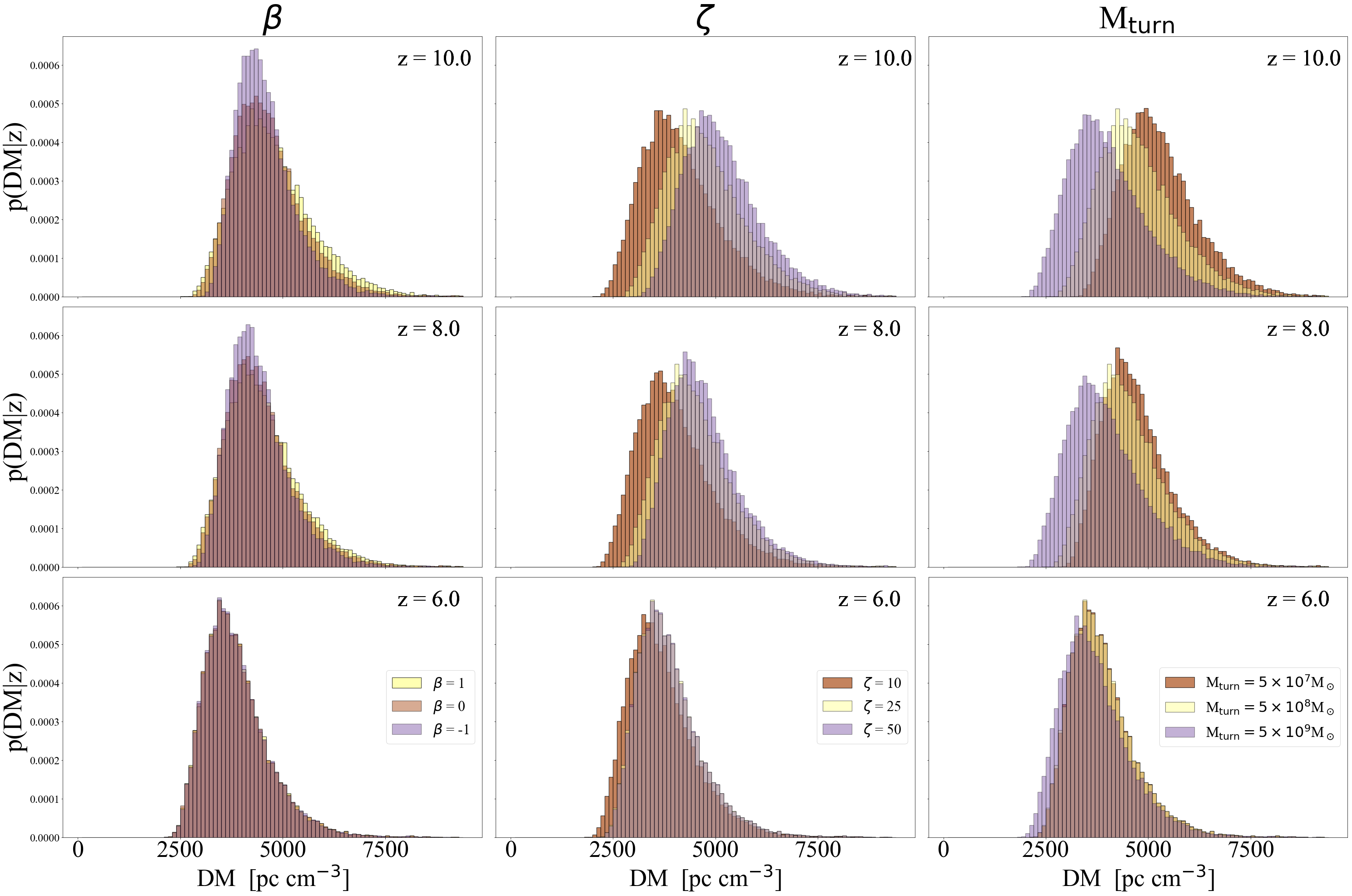}
  \caption{Evolution of the individual sightline DM probability distributions for a variety of reionization scenarios encapsulated by the density-ionization parameter $\beta$, ionizing efficiency $\zeta$ and mass scale of the ionizing sources $M_{\rm{turn}}$. Notice how the different reionization scenarios begin to distinguish themselves at higher redshifts. In each panel, the gold distribution corresponds to the fiducial reionization scenario of $\beta = 1$, $\zeta = 25$, $M_{\rm{turn}} = 5\times 10^8$M$_\odot$ and $R_{\rm{mfp}} = 30$Mpc.  }
\label{fig:Prob_DM_big_3}
\end{figure*}

We consider the distribution of the individual DM sightlines as a function of the astrophysics and morphology of the EoR. FRBs observed at low $z$ are more likely to have low DM sightlines due to less chance of interactions with free electrons in the IGM. From Figure \ref{fig:Prob_DM_single} we see that the resulting DM probability distribution is highly non-Gaussian and skewed to low DM. Note that the contaminants $\overline{\rm{DM}}_{\rm CGM}$ and $\overline{\rm{DM}}_{\rm ISM}$ can produce high DM fluctuations, even at low redshift. Removal of these contaminated sightlines are required in order to make precise deductions about the state of the IGM using FRB DM statistics. This might be especially difficult to do for high redshift FRBs since FRBs observed at high redshift are more likely to interact with free electrons from the IGM and so large DM sightlines become more likely (see Figure \ref{fig:Prob_DM_single}). At higher redshifts, the distribution functions tend to become better approximated as Gaussian. Previous studies such as \cite{Yoshiura_Shintaro}, use the variance, $\sigma^2 = \langle(\rm{DM} - \overline{\rm{DM}})^2 \rangle$, of the DM distributions to estimate the maximum size of the ionized regions at each $z$. Since the maximum size of the ionized regions depends on the astrophysical parameters, our approach is complimentary. The astrophysics driving the EoR will determine the evolution of the $\rm{DM}(z)$ probability distributions. For example, scenarios with larger $\zeta$ or smaller $M_{\rm{turn}}$ tend to have DM distributions skewed to higher DM since reionization begins early, which increases the relative likelihood of finding high DM sightlines by increasing the likelihood of interaction with free electrons. Meanwhile, scenarios where the EoR unfolds as inside-out, tend to have high density regions in $\delta$ couple to regions of high fraction of ionized hydrogen, i.e. the product $ \overline{x_{\rm{HII}} \delta}  \sim n_e$ is larger than the corresponding outside-in scenario where the high density regions couple to low fraction of ionized hydrogen. In this scenario, the free electron regions tend to be denser in inside-out models than the corresponding outside-in models. As a result, scenarios where reionization unfolds with $\beta > 0$ increases the likelihood of high DM sightlines. This is reflected in the DM$(z)$ distributions in Figure \ref{fig:Prob_DM_big_3} where there is a larger portion of distribution in the high DM portion of the distribution compared to outside-in maps where the distribution is skewed to lower DMs. 

These parameters influence the shape of the DM distribution as well as their evolution in redshift. Since $\overline{\rm{DM}}$ is derived from these DM distributions, then the underlying astrophysics and morphology of the EoR can be detected directly from $\overline{\rm{DM}}$. In the following Sections, we build our intuition on how the astrophysics and morphology of the EoR affect $\overline{\rm{DM}}$.

%The astrophysics driving the EoR are imprinted onto the ionization field $x_{\rm{HII}}$. High redshift FRBs detected during the EoR will have these astrophysical properties imprinted the integrated profile of $n_e$.

\begin{figure*}
  \includegraphics[width=0.99\textwidth]{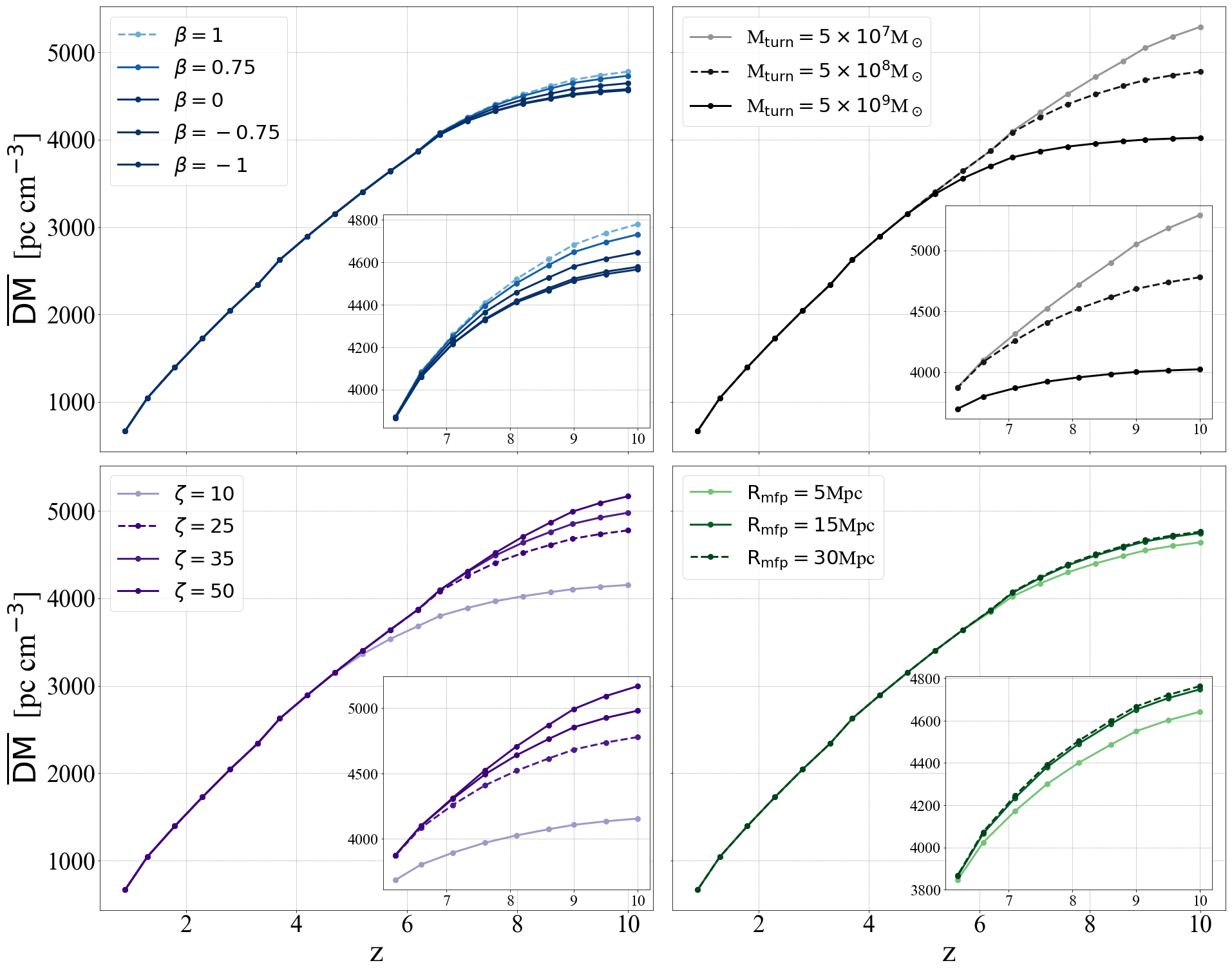}
  \caption{$\overline{\rm{DM}}$ for a variety of density-ionization correlations $\beta$ (upper left), mass scale of the ionizing sources $M_{\rm{turn}}$ (upper right), ionizing efficiency  $\zeta$ (lower left), and mean free path $R_{\rm{mfp}}$ of the ionizing photons. Notice how high ionizing efficiency of the sources and smaller masses of the ionizing sources lead to an early onset reionization, and so an increase in $\overline{\rm{DM}}$ at that redshift. Inside-out reionization models $\beta > 0$, lead to an increase in $\overline{\rm{DM}}$, since the free electron number density in ionized regions is greater than the corresponding ionized region in outside-in models $\beta < 0$. In each panel the dotted curve corresponds to same reionization scenario $\beta = 1$, $\zeta = 25$, $M_{\rm{turn}} = 5\times 10^8$M$_\odot$ and $R_{\rm{mfp}} = 30$Mpc. We use this fiducial reionization scenario in our forecasts in Section \ref{sec:forecasts}. }
\label{fig:DM_z_theory}
\end{figure*}

 \subsection{Astrophysical Signature on $\overline{\rm{DM}}$}
 \label{subsec:AstroSig}
  Local fluctuations in $n_e$ make it difficult to deduce the astrophysics from individual sightlines. Instead we average over all sightlines to remove these fluctuations. In doing so, we can predict the signature of the astrophysical parameters on $\overline{\rm{DM}}$ in Equation \ref{eq:DM_obs_z_mean_final}. Since the presence of neutral hydrogen in the IGM causes a flattening of the  $\overline{\rm{DM}}$ curve at the onset of neutral hydrogen, then the astrophysical parameters, which determine the timing of this flattening, can be deduced from $\overline{\rm{DM}}$. For example, the ionizing efficiency $\zeta$ increases the output of UV photons from the ionizing sources, which for larger values of $\zeta$, results in shifting the onset of reionization to higher redshifts.  In this scenario, the IGM is ionized earlier and the flattening of the $\overline{\rm{DM}}$ curve occurs at larger $z$. Conversely, decreasing the ionizing efficiency of the sources shifts the flattening of the $\overline{\rm{DM}}$ curve to lower redshifts. Therefore, if we study the dependence of $\overline{\rm{DM}}$ on $\zeta$ at fixed $z$ (within the EoR), increasing the ionizing efficiency will increase the mean DM of the FRBs at that redshift. We find a similar dependence for $\overline{\rm{DM}}$ on $M_{\rm{turn}}$. This is the mass scale for a source to begin efficiently producing UV photons which similarly alters the onset of reionization. Lower values of $M_{\rm{turn}}$ allow the EoR to start early, which shifts the flattening of $\overline{\rm{DM}}$ to higher redshifts, while larger values of $M_\odot$, delays reionization, pushing the flattening of $\overline{\rm{DM}}$ to lower redshifts. We find that $\overline{\rm{DM}}$ is less sensitive to $R_{\rm mfp}$ as compared to the other EoR parameters. Once $R_{\rm mfp}$ is increased beyond the size that is physically possible at given $z$,  $\overline{\rm{DM}}$ loses all sensitivity to the parameter. 
  
  In general, these astrophysical parameters determine the mean ionization fraction $\overline{x}_{\rm{HII}}$ at each redshift $z$, which $\overline{\rm{DM}}$ depends on. One could approach the study of $\overline{\rm{DM}}$ on $\overline{x}_{\rm{HII}}$ by adopting a model for the evolution of $\overline{x}_{\rm{HII}}$ on $z$ without invoking the dependence of astrophysical parameters. However, these parameters can also have a secondary affect on $\overline{\rm{DM}}$ through the cross term in Equation \ref{eq:DM_obs_z_mean_final}. For example, if the Universe reionizes with turnover masses $M_{\rm{turn}} ~ 10^{10}$M$_\odot$, then the size of the ionized regions are larger compared to a scenario with smaller turnover which increases the cross term in Equation \ref{eq:DM_obs_z_mean_final}. Physically this means that there are more free electrons for the FRBs to interact with. Increasing the maximum size of the ionized regions $R_{\rm mfp}$ will maximize the interaction between FRBs and free electrons for given EoR model with fixed $\zeta$ and $M_\odot$. This maximises $\overline{\rm{DM}}$.
  
 %In general, the astrophysical parameters determine the mean ionization fraction $\overline{x}_{\rm{HII}}$ at each redshift $z$ which $\overline{\rm{DM}}$ depends on. Alternatively, one could approach the study of $\overline{\rm{DM}}$ on $\overline{x}_{\rm{HII}}$ by adopting a model for the evolution of $\overline{x}_{\rm{HII}}$ on $z$ without invoking the dependence of astrophysical parameters. However 
 
 The sensitivity of $\overline{\rm{DM}}$ to $\zeta$, $M_{\rm{turn}}$ and $R_{\rm mfp}$ increases as we observe to FRBs at higher redshifts. This is due to the FRBs having interacted with the ionization history of the universe for longer and so Equation \ref{eq:DM_obs_z_mean_final} carries more information about the EoR. Conversely,  $\overline{\rm{DM}}$ loses all sensitivity to the astrophysics of the EoR as the entire Universe is reionized, referring to Figures \ref{fig:DM_z_theory}, all models converge at $z = 6$ which in our models correspond to an entirely ionized IGM.

\subsection{Morphological Signature on $\overline{\rm{DM}}$}
\label{subsec:MorphSig}
%From Equation \ref{eq:DM_obs_z_mean_final}, the mean DM of high redshift FRBs is sensitive to the density-ionization product $ \delta x_{\rm HII}$. The method in which $x_{\rm HII}$ couples to the underlying density field $\delta$ will have consequences for $\overline{\rm{DM}}$. Inside-out scenarios, i.e. scenarios where $ \delta$ is positively correlated with $x_{\rm HII}$ ($\beta > 0$), and so high density regions couple to high ionized fractions in $x_{\rm HII}$ denser regions of free electrons. For the outside-in scenario, i.e. $\beta < 0 $, the density field is negatively correlated with the ionization field, and so the underdense regions in $\delta$ correspond to high fractions of ionized hydrogen. As a result we expect the ionized bubbled to have a lower In this scenario, we expect the ionized regions to be dense,  leading to larger values of $n_e$ compared to the identical ionized regions in outside-in models. 

From Equation \ref{eq:DM_obs_z_mean_final}, the mean DM of high redshift FRBs is sensitive to the density-ionization product $ \delta x_{\rm HII}$. The method in which $x_{\rm HII}$ couples to the underlying density field $\delta$, will have consequences for the $\overline{\rm{DM}}$. Inside-out scenarios, i.e. scenarios where $\beta > 0$ (positive correlation between $\delta$ and $x_{\rm{HII}}$), high density regions couple to high ionized fractions in $x_{\rm{HII}}$. This  results in the ionized regions being denser in free electrons, leading to an increase in $\overline{\rm{DM}}$ compared to other morphologies. For example,  the outside-in scenario, where $\delta$ and $x_{\rm{HII}}$ are negatively correlated ($\beta < 0 $), the underdense regions in $\delta$ correspond to high fractions of ionized hydrogen. As a result, the free electron density within the ionized regions are comparatively smaller. Referring to Figure \ref{fig:DM_z_theory}, we see that inside-out morphologies lead to an increase in the mean DM of high redshift FRBs as compared to outside-in models. Intermediate values of $\beta$ can be interpreted as follows; as $\beta$ is increased from the uncorrelated scenario, $\beta = 0$, (where the ionized regions are random with respect to $\delta$ ) to $\beta = 1$, the high density regions in $\delta$ becoming increasingly likely to couple to ionized regions in $x_{\rm{HII}}$. The mean $n_e$ within bubbles monotonically increases until $\beta = 1$ where all high density regions correspond to ionized bubbles and $\overline{\rm{DM}}$ is maximized with respect to $\beta$. Conversely, as we decrease $\beta$ from $\beta = 0$ to $\beta = - 1$, the high density regions increasingly couple to regions of low ionized fraction in $x_{\rm{HII}}$ which monotonically decreases the mean $n_e$ of the ionized regions. As a result, the product $n_e \sim \delta x_{\rm HII}$ is decreased, which leads to a decrease in the mean DM of these models. As a result, inside-out scenarios receive a boost in average DM due to the increase of $\overline{n}_e$ compared to outside-in driven models. The morphological signature on $\overline{\rm{DM}}$ is different that the astrophysical parameters since the morphology directly influences the mean density of free electrons, $n_e$ within the ionized bubbles without changing the timing of reionization. The contrast in $\overline{\rm{DM}}$ between the extreme morphologies is greatest for FRBs observed at highest redshift. The longer the exposure of the FRB to the ionization history, the more sensitive $\overline{\rm{DM}}$ will be to the morphology. Conversely, as we observe FRBs at lower redshifts, there hasn't been enough exposure to the EoR morphology to distinguish between different $\beta$ models. Therefore $\overline{\rm{DM}}$ loses all sensitivity to $\beta$ as $\overline{x}_{\rm{HII}} \rightarrow 1$. In Section \ref{sec:Results}, we determine the number of FRBs required to make a measurement of $\overline{\rm{DM}}$ precise enough to place constraints on $\beta$ as well as the astrophysical parameters.

In the following Section, we use generate mock data by sampling the fiducial DM distributions at each redshift given our choice of fiducial EoR and morphological parameters. In Section \ref{sec:MCMC_setup}, we forecast the type of constraints that can be placed on $\beta$ as well as the remaining EoR parameters through measurement of $\overline{\rm{DM}}$.

\section{Forecasts}
\label{sec:forecasts}

In this Section we use the formalism of Section \ref{sec:models} to forecast the constraints that can be placed on the EoR through measurement of high redshift FRB DMs. Since high redshift FRBs have not yet been detected, we simulate a mock observation of high DM FRBs  under a fiducial reionization scenario. It should be noted that it is assumed that all generated FRBs are observed with accompanied redshift localization where we take the uncertainty on the redshift, $\sigma_z = 0$. This may seem an ambitious assumption, but \cite{Walters2018} notes that with a mid- to large-size optical survey, it should be feasible to obtain about 10 redshifts for host galaxies per night. In this Section, we outline our model for generating this mock observation as well as discuss our fiducial reionization scenario.  In Section \ref{sec:Results}, we present the results of these forecasts.

%We have not yet made any detections of the EoR and so in order to make a mock observation of high redshift FRBs, we must pick a fiducial reionization scenario. Our fiducial EoR model is produced by fixing the astrophysical and morphological parameter $\beta$ from Section $\ref{sec:models}$. We choose EoR parameters $\zeta_0 = 25$ , M$_{\rm{turn},0} = 5\times10^8 $M$_\odot$, and R$_{\rm{mfp},0} = 30$Mpc as well as $\beta = 1$. The astrophysical parameters are consistent with previous studies such as \cite{kSz}, while the morphological parameter, $\beta = 1$,  corresponds to an inside-out reionization scenario. These parameters determine the probability distribution of DMs at each redshift. We sample individual DMs from these distributions corresponding to detection of an FRB at that redshift with the selected DM.  

\subsection{Intrinsic FRB Statistics}
\label{sec:intrinsic_stats}
Since FRBs observed after the EoR do not contain any information about the ionization history of the Universe, only high redshift FRBs observed during the EoR contribute to our forecasts. FRBs at these redshifts have not yet been observed and may be rare. To get a more realistic sense of how many intrinsic FRBs that can potentially be observed given a capable high DM experiment, we use an existing theoretical model of source count distributions of FRBs at each DM. From this theoretically motivated count of FRBs within $z > z_{\rm{EoR}}$, we can populate our mock catalogue. We first define an intrinsic source count distribution of FRBs. It will be from this distribution that we populate the redshift bins of our fiducial sample for our forecasts. For this, we choose a source count distribution that traces the star formation rate (SFR) \citep{Liam}. While other source count distributions have been proposed, \cite{Niino} shows that the the density of FRBs ($\rho_{\rm{FRB}}$) increases with redshift, closely resembling cosmic star formation history. We follow this prescription and use the following simple top-heavy distribution for the number of FRBs per DM,

\begin{equation}\label{eq:source-count}
    \frac{dn}{d\rm{DM}} = \frac{\rho_{\rm{FRB}}(z)}{(1+z)}\frac{dV}{dz}\frac{dz}{d\rm{DM}}
\end{equation}

where, $\frac{dV}{dz}$ is the comoving volume element.

Tracing the cosmic star formation history, we take the density to proportional to the SFR density \citep{Madau+lots,Madau1998},

\begin{equation}
    \rho_{\rm{FRB}}(z) \propto \rho_{\rm{SFR}}(z) = 0.015 \frac{(1+z)^{2.7}}{1+((1+z)/2.9)^{5.6}} \textrm{ M}_\odot \textrm{ yr}^{-1} \textrm{ Mpc}^{-3}
\end{equation}

 What relies on the model in this source count distribution is the $\frac{dz}{d\rm{DM}}$ factor. As mentioned in Sections \ref{subsec:AstroSig} and \ref{subsec:MorphSig}, the DM--z relation is sensitive to reionization parameters. Currently, the widely used DM–z relation is linear
 
 \begin{equation}\label{eq:dm-z}
    \rm{DM}(z) = C\times z \textrm{ pc cm}^3
\end{equation}

where C is often taken to be 1000 \citep{Niino} or 1200 \citep{Ioka}. These linear relations approximate the redshift to an accuracy of about 2\% for $z<2$ \citep{Petroff2016}. As shown in Figure \ref{fig:DM_z_theory}, the model of reionization affects the shape of the DM--z relation, especially at high redshift. In order to place constraints on reionization, high redshift samples are paramount. Therefore, in order to compute the source count distribution for the fiducial model, we calculate $\frac{dz}{d\rm{DM}}$ by taking numerical derivatives of the corresponding fiducial DM--z curve. Measurements of the EoR parameters have not yet been made, so in order to produce a mock sample of observed high redshift FRBs, we must assume a fiducial reionization scenario. Our fiducial EoR model is produced by fixing the astrophysical and morphological parameter $\beta$ from Section $\ref{sec:models}$. We choose EoR parameters $\zeta_0 = 25$ , M$_{\rm{turn},0} = 5\times10^8 $M$_\odot$, and R$_{\rm{mfp},0} = 30$Mpc as well as $\beta = 1$. The astrophysical parameters are consistent with previous studies such as \cite{kSz}, while the morphological parameter, $\beta = 1$,  corresponds to an inside-out reionization scenario. 
%These parameters determine the probability distribution of DMs at each redshift. 
%We sample individual DMs from these distributions corresponding to detection of an FRB at that redshift with the selected DM.  
%that line above is talking about how the Fiducial model fixes the DM distribution, should
%probably put that part later in this paragraph where you're actually describing the sampling

%look down

 The fiducial DM--z curve is the light blue dashed line ($\beta = 1$) shown in the top left panel of Figure \ref{fig:DM_z_theory}. Now that the CDF is defined, we can build our mock data set.

 %IS THIS THE SAMPLING YOU MEAN?? LIKE FROM THE HISTOGRAMS where we assign the DMs to the number of DMs based on their redshift bin???

%yeah sampling as in we're drawing a particular DM for each z from the DM histograms.
%maybe sampling is the wrong word? "Drawing" N DMs from the distributions at each z
%where N is determined by the source count distribution

% for sure got it, I'll elaborate on this then in this paragraph 

\subsection{Mock Catalogue of FRBs}
\label{sec:mock_catalogue}

We build our sample of FRBs using inverse transform sampling whereby a given number of random samples is drawn from a probability distribution given its CDF. This populates each redshift bin, of bin width $\delta_z$ = 1, with FRBs according to the CDF. The method is the following, where our random variable $X$ is the FRB source count: 

\begin{enumerate}
    \item Define a random variable, $X$, whose distribution is described by the CDF, $F_X$.
    \item Generate a random number $u$ from a uniform distribution in the interval $[0,1]$. This number will be interpreted as a probability. 
    \item Compute the inverse of of the CDF, that is $F^{-1}_X(u)$.
    \item Compute $X = F^{-1}_X(u)$. Now the random variable $X$ with distribution $F_X$ has been generated. 
\end{enumerate}

Once we build our sample of FRBs, the DMs for each source are drawn from the probability distribution, $p(\rm{DM}|\rm{z})$, associated with the fiducial model (see Figure \ref{fig:Prob_DM_single}). This produces line of sight fluctuations, ensuring that every FRB has a unique DM, even when in the same redshift bin. It may be noted that we only account for fluctuations in the DM distribution of FRBs. The spacial distribution of FRBs is not accounted for here. In actuality, the spacial distribution of FRBs will be positively correlated with the underlying matter distribution. We find that the contribution of the host bubble, or lack there of, to the total DM from the line of sight during reionization is negligible and we proceed without populating halos with sources. At this point, we are ready to move on to performing the MCMC on the sample.

\subsection{MCMC setup}
\label{sec:MCMC_setup}

%FRBs observed after the EoR do not contain any information about the state of the IGM. As a result only FRBs observed during the EoR can contribute to the constraints placed. FRBs in this regime have not yet been observed and may be rare. To get a more realistic sense of how many intrinsic FRBs that can potentially be observed given a capable high DM experiment, we use an existing theoretical model of source count distributions of FRBs at each DM. From this theoretically motivated count of FRBs within $z > z_{\rm{EoR}}$, we to pick the number of FRBs available to use in our mock catalogue. We consider three scenarios (i)optimistic, (ii) moderate, (iii) pessismistic, where each scenario relates to the expected number of detectable FRBs that can be observed during reionization. We present the results of such a forecast in Section \ref{sec:Results}.

We place the mean of the individual sightline $\rm{DM}$s of our mock FRB catalogue into a vector $\mathbf{\overline{\rm{DM}}_S}$ corresponding to the mean of the sample FRBs for each redshift $z$. For such a measurement, the uncertainties on $\mathbf{\overline{\rm{DM}}_S}$ are the sum of the instrumental systematic errors in measuring the individual sightline DMs and the uncertainties in $\overline{\rm{DM}}_S$ due to sample variance. The instrument errors on the individual DM are assumed to be small and so we do not model the instrumental errors and only include the errors due to sample variance. The uncertainties due to sample variances on $\overline{\rm{DM}}_S$ are

\begin{equation}
    \label{eq:sample_variance}
    \sigma = \frac{s_{N-1}}{\sqrt{N}}
    %\sigma^2 = \frac{1}{N - 1}\mathlarger{\mathlarger{‎‎\sum}}_{i=0}^{N}
\end{equation}
where $N$ are the number of FRBs comprising the sample and $s_{N -1 }$ is the measured sample variance given by 
\begin{equation}\label{eq:sample_variance2}
    s^2_{N-1} = \frac{1}{N - 1}{\sum_{i=0}^{N}}(\rm{DM}_i - \overline{\rm{DM}}_S)^2
\end{equation}
where $\rm{DM}_i$ are the individual sightline DMs sampled from the probability density functions generated by our fiducial model, and $\overline{\rm{DM}}$ is the mean of such a sample. Our forecasts consider different cases of $\sigma_S$ by considering different total number $N$ of FRBs observed. To place constraints on the on the EoR parameters $\boldsymbol \theta = \beta, \zeta, M_{\rm turn}, R_{\rm{mfp}}$, we evaluate the probability of $\mathbf{\theta}$ given measurement of the mean DM from the samples from our fiducial EoR model defined in Section \ref{sec:mock_catalogue}. This is the posterior $p(\boldsymbol  \theta |  \overline{\rm{DM}}_S)$. We can evaluate the posterior $p(\boldsymbol \theta | \mathbf{d_{\rm S}})$ through Bayes theorem:
\begin{equation}
\label{eq:bayes}
p(\boldsymbol \theta | \overline{\rm{DM}}_S) \propto p( \overline{\rm{DM}}_S| \boldsymbol \theta ) p(\boldsymbol \theta), 
\end{equation}
where $p( \overline{\rm{DM}}_S| \boldsymbol \theta )$ is the likelihood function and $p(\boldsymbol \theta)$ is the prior on the EoR parameters $\boldsymbol \theta$. Since the likelihood function is non-analytic in the EoR parameters $\boldsymbol \theta$, we use \texttt{21cmFAST} to generate a model density and ionization field representative of the IGM with parameters $\beta$, $\zeta$, $M_{\rm{turn}}$, $R_{\rm{mfp}}$. To generate the density and ionization field with the morphology indicative of the model $\beta$, we use the same procedure described in \cite{me!}. From this model reionization and density field, we generate a lightcone for each line of sight, and evaluate $\rm{DM}$ for each of these lines of sight. We then average all sightlines together to evaluate $\overline{\rm{DM}}$ for this reionization model. The mean DM of all sightlines for this model is compared to the fiducial mean DM of the mock FRBs through the $\chi^2$ statistic. The likelihood $p( \mathbf{d}_{\rm{S}} | \boldsymbol \theta )$ is then computed as: 

\begin{equation}
\label{eq:likelihood}
    p( \mathbf{\overline{\rm{DM}}_S}| \boldsymbol \theta ) \propto \exp \left[-\frac{1}{2} \sum_{z} \frac{ \left( \overline{\rm{DM}}_{\rm{model}} - \mathbf{\overline{\rm{DM}}_S} \right)^2}{\sigma^2_S} \right] ,
\end{equation} 
%The non-Gaussianity is evident in the probability distributions of Figure \ref{fig:Prob_DM_big_3} in , and so this approach serves as an approximation to the true likelihood capturing only the Gaussian information contained in $\mathbf{\overline{\rm{DM}}_S}$.
%Our sample data is generated from the non-Gaussian distributions in Section \ref{sec:intrinsic_stats}. Since the exact form of the non-Gaussianity in $p(\mathbf{\overline{\rm{DM}}_S} | z)$ depends on the precise reionization scenario, we assume the experimenter taking observations on $\overline{\rm{DM}}$ does not know apriori the form of the non-Gaussian error bars. Future studies who model the true non-Gaussian elements of the posterior can take advantage of the additional information contained in $\mathbf{\overline{\rm{DM}}_S}$. Our approach is therefore a conservative one, and only captures the Gaussian information contained in $\mathbf{\overline{\rm{DM}}_S}$. The form of the non-Gaussianity is highly depdendent on how the EoR actually unfolds, so give measurements of $\mathbf{\overline{\rm{DM}}_S}$ at each $z$, we can know exactly what the form of the non-Gaussianity without fitting to that too. We leave that to future work.
where we have assumed the errors on $\mathbf{\overline{\rm{DM}}_S}$ to be Gaussian and independent. The Gaussianity of the likelihood is a valid assumption since for larger samples of $\rm{DM}$, the mean $\overline{\rm{DM}}$, of these samples tend to be Gaussian distributed according to the central limit theorem. However since there are indeed correlations between redshift bins, the independence of the likelihood in terms of $z$ serves as an approximation. We consider the mean DM of FRBs measured from redshifts $z = 8$ to $z  = 10$ in steps of $\Delta z = 1$ corresponding to the redshifts that contain the largest sensitivity to the EoR parameters.  Inclusion of more redshifts do not significantly alter our conclusions and so for computational simplicity we exclude them from our forecasts. We place uniform priors on each of the EoR parameters $\boldsymbol \theta$ within $p(\boldsymbol \theta)$. Since $\beta$ is only defined from $-1 \le \beta \le 1$, we place the prior  $-1 \le \beta \le 1$ which covers the entire possible physical range of EoR morphologies.  For $R_{\rm mfp}$ we use $5\,\textrm{Mpc} < R_{ \rm mfp} < 160\,\textrm{Mpc}$ which spans the all possible sizes consistent with the length of our simulation boxes. For $\zeta$, we place the range $5 < \zeta < 100$ which encapsulates the entire physically allowed duration of reionization histories \cite{kSz}. Finally for $M_{\rm turn}$, we use values of $10^7 M_\odot <  M_{\rm turn} <  10^{10} M_\odot$, which are physically motivated by the atomic cooling threshold and by constraints on the faint end of UV luminosity functions \citep{Park}. Using the sampling discussed in Section \ref{sec:intrinsic_stats}, we generate mock data and fit to them via the likelihood 

\begin{equation}
\label{eq:likelihood_2}
    p( \mathbf{\overline{\rm{DM}}_{S_i}}| \boldsymbol \theta ) \propto \prod_i \exp \left[-\frac{1}{2} \sum_{z} \frac{ \left( \overline{\rm{DM}}_{\rm{model}} - \mathbf{\overline{\rm{DM}}_{S_i}} \right)^2}{\sigma^2_{S_i}} \right] .
\end{equation} 
To sample our posterior distribution, we use a Markov Chain Monte Carlo (MCMC) approach, as implemented by the affine invariant MCMC package \texttt{emcee} \citep{emcee}.

\section{Results}
\label{sec:Results}
Here we present the MCMC results of our forecast discussed in Section \ref{sec:MCMC_setup} corresponding to measurement of $N$ high redshift FRBs observed between $z = 8$ to $z = 10$, and distributed in $z$ according to the CDF described in Section \ref{sec:intrinsic_stats}. We repeat this mock observation for three different total number of measured FRBs. We use $N = 10^2, 10^4, 10^5$, where these observed FRB counts span a reasonable range of sample variances. The fiducial reionization scenario has parameters $\beta = 1$, $\zeta = 25$ , M$_{\rm{turn}} = 5 \times 10^8$M$_\odot$ and $R_{\rm{mfp}} = 30$Mpc.

%Here we present the MCMC results of our forecast discussed in Section \ref{sec:MCMC_setup}. We separate our forecast into three observational scenarios:
%$N$ high redshift FRBs are observed between $z = 6$ to $z = 10$ and distributed in $z$ according to the CDF described in Section \ref{sec:intrinsic_stats}. We repeat this mock observation for two different total number of measured FRBs. We use $N = 100, 600, 10^4$, where these observed FRB counts span a reasonable range of sample variances. The fiducial reionization scenario has parameters $\beta = 1$, $\zeta = 25$ , M$_{\rm{turn}} = 5 \times 10^8$M$_\odot$ and $R_{\rm{mfp}} = 30$Mpc.
%Measurement of 10 FRBs in each redshift bin between $z = 6$ to $z = 10$. Should instruments be able to detect high redshift FRBs, this scenario is meant to be a more accurate representation of the kind of constraints that can be placed on the EoR using these early limits. We use the same fiducial reionization scenario as (i). 

\begin{figure}
  \includegraphics[width=0.49\textwidth]{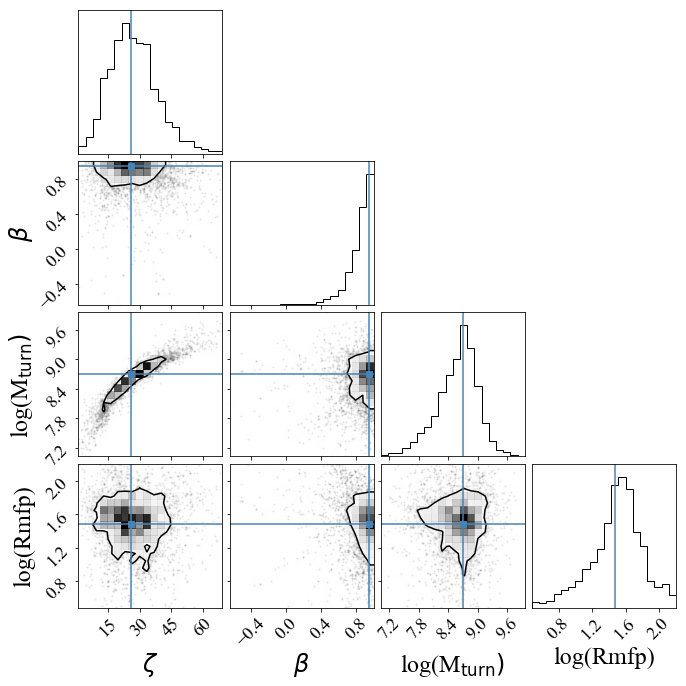}
  \caption{Posterior distributions for measurement of $\overline{\rm{DM}}$ for $10^4$ FRBs distributed between redshifts $ 8 \leq z \leq 10$ according to the source count distribution in Section \ref{sec:intrinsic_stats}. The 68$\%$ credibility regions are shown. Such a measurement can rule out uncorrelated $\beta = 0$ and outside-in reionization $\beta < 0$ at 68$\%$CR. }
\label{fig:MCMC_N600}
\end{figure}

\begin{figure}
  \includegraphics[width=0.49\textwidth]{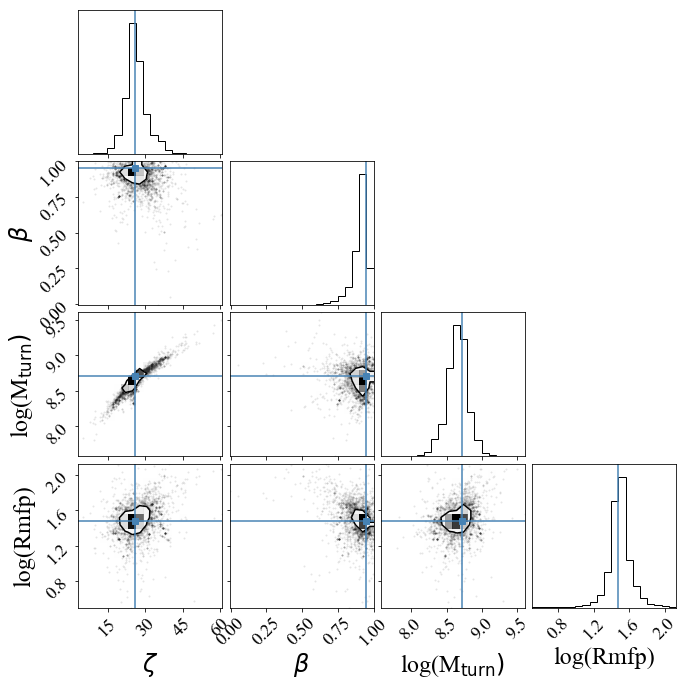}
  \caption{Posterior distributions for measurement of $\overline{\rm{DM}}$ for $10^5$ FRBs distributed between redshifts $ 8 \leq z \leq 10$ according to the source count distribution in Section \ref{sec:intrinsic_stats}. Using such a measurement, we can rule out uncorrelated and outside-in reionization scenarios at $68\%$CR.}
\label{fig:MCMC_N10000}
\end{figure}

\subsection{Larger Sample Sizes}
\label{sec:N600_z8-10} 
In this scenario we detect $N$ FRBs, distributed across the redshift bins $z = 8 - 10$ according to the theoretically motivated source count distribution discussed in Section \ref{sec:intrinsic_stats}. Figure \ref{fig:MCMC_N600} and \ref{fig:MCMC_N10000} show the results of this forecast for cases corresponding to $N = 10^4$ and $N = 10^5$ respectively. From the posterior of both Figures, we see that there are clear degeneracies between $\zeta$ and $M_{\rm{turn}}$. This is due to both parameters establishing the redshift in which the flattening of  $\overline{\rm{DM}}$ occurs. This degeneracy is pronounced in Figure \ref{fig:Prob_DM_big_3} where changing the values of $\zeta$ and $M_{\rm{turn}}$ result in translating the distribution along the horizontal axis. By examining the $68\%$ credibility regions (CR) in Figure \ref{fig:MCMC_N600}, we can see that measurement of $N = 10^4$ FRBs within these redshift bins can constrain $\zeta$ to $\zeta = 25.5^{+11.5}_{-10.5}$ and $\textrm{log}(\rm{M}_{\rm{turn}}) = 8.65^{+0.29}_{-0.49}$. By placing constraints on these parameters (the ionizing efficiency and the halo mass scale of the UV sources), one can place constraints on the timing and duration of reionization. We find that with $N = 10^4$ FRBs in these redshift ranges, we can constrain the duration, $\Delta z$, of reionization (duration between $0.25 \leq \overline{\rm{x}}_{\rm{HII}} \leq 0.75$) to $\Delta z = 2.1^{+0.50}_{-0.30}$, and the midpoint $z_{\rm{mid}} = 7.8^{+0.20}_{-0.20}$, at $68\%$ credibility. Referring to the posterior in Figure \ref{fig:MCMC_N10000}, the constraints on $\zeta$ and $M_{\rm{turn}}$ are tighter for the extreme case of $N=10^5$ FRBs where we can constrain $\zeta$ to within $\zeta = 25^{+7}_{-9}$ at $95\%$CR and log$(\rm{M}_{\rm{turn}}) = 8.76^{+0.14}_{-0.46}$ at $95\%$CR. With constraints on these parameters we can place constraints on the duration of reionization, $\Delta z = 2.0^{+0.5}_{-0.4}$, at $\%95$CR and the midpoint of reionization, $z = 7.8^{+0.4}_{-0.2}$ at 95$\%$CR. 

The correlation parameter $\beta$ does not share degeneracies with these parameters since it does not affect the timing of reionization, rather it affects the mean density, $\overline{n}_e$, of free electrons in the ionized region. We find from the posterior that measurement of $10^4$ FRBs can distinguish between the sign of $\beta$. Since the sign of $\beta$ corresponds to the type of correlation between $\delta$ and $x_{\rm{HII}}$, we find that measurement of $\overline{\rm{DM}}$ using $10^4$ FRBs can rule out $\beta < 0$ (outside-in) scenarios and $\beta = 0$ (uncorrelated scenarios) at 95$\%$ CR. In the more extreme case of $N=10^5$ FRBs, we can further rule out uncorrelated and outside-in scenarios at $99\%$ CR. Measurement of $10^4$ FRBs between $8 < z < 10$ is sufficient to constrain the order of magnitude of $R_{ \rm mfp}$ at $68\%$CR. For the case of $N=10^5$ FRBs, our models can constrain the order of magnitude of $R_{ \rm mfp}$ at $95\%$CR. 

%discern the small fluctuations in $\overline{\rm{DM}}$ due to $R_{ \rm mfp}$. In this scenario we can place constraints on the order of magnitude on the maximum size of the ionized regions.

%The posterior demonstrates that our models are less sensitive to $R_{ \rm mfp}$ as compared to the other parameters. This is unsurprising, since the variation of $R_{ \rm mfp}$ produces fluctuations of $\overline{\rm{DM}}$ well within the sample variance errorbars in Section \ref{sec:MCMC_setup}.
%Note that using a  Gaussian model of the likelihood, we find that we cannot place 

\subsection{Smaller Sample Sizes}
\label{sec:10uniform_z6-10} 
\begin{figure}
  \includegraphics[width=0.49\textwidth]{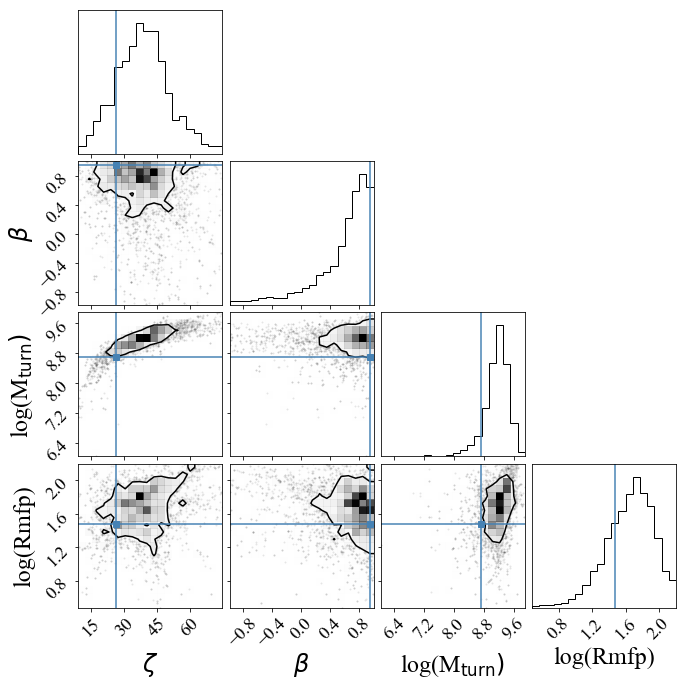}
  \caption{Posterior distributions for measurement of $\overline{\rm{DM}}$ for $100$ FRBs distributed between $ 8 \leq z \leq 10$ according to the source count distribution in Section \ref{sec:intrinsic_stats}. The 68$\%$ credibility regions of our measurements are shown. This measurement can rule out extreme EoR models, for example, scenarios where the Universe is ionized by $z = 8$.}
\label{fig:MCMC_N100}
\end{figure}

In this scenario we measure $100$ FRBs distributed across redshift bins between $8 \le z \le 10$, again using the source count distribution outlined in section \ref{sec:intrinsic_stats}.  We show the posterior of such a measurement in Figure \ref{fig:MCMC_N100}. Our interpretation of the degeneracy between the parameters is identical to \ref{sec:N600_z8-10}. We see from the posterior of Figure \ref{fig:MCMC_N100} that smaller samples of FRBs lead to biased fits due to cosmic variance. However even with such small sample sizes, $68\%$ of the contours lie within $\beta > 0$ suggesting that we can still rule out both uncorrelated and outside-in reionization scenarios at $68\%$CR. We see from the posterior that we can rule out models with $\zeta$ and M$_{\rm{turn}}$ outside the range $23 \leq \zeta \leq 55$ and $4 \times 10^9$M$_\odot \leq M_{\rm{turn}} \leq 3 \times 10^9$M$_\odot$ at $68\%$CR. Ruling out this region of parameter space is tantamount to setting broad constraints on the timeline of reionization. For example, this region excludes scenarios where the Universe is still neutral at redshift $z = 10$, which would severely flatten $\overline{\rm{DM}}(z)$ between $8 \le z \le 10$. We can rule these models out at $68\%$CR. Similarly this region excludes models where the Universe is more than $60\%$ ionized by redshift $z = 8$, which would reduce the flattening of $\overline{\rm{DM}}$ between $8 \le z \le 10$. We can rule out these scenarios at $68\%$CR.

%and larger values of $M_{\rm{turn}}$ are preferred. To understand why this is, consider the DM probability distributions, $p(DM | z)$ in Section \ref{subsec:DM_Distributions}. These distributions are non-Gaussian and the most probable DM value is generally not equal to $\overline{\rm{DM}}$. For such small sample sizes ($<10$ FRBs in each redshift bin), when drawing our mock DM samples from $p(DM | z)$, we mostly obtain DM samples that underestimate the true mean DM at that $z$. As a result, our mock measurement of $\overline{\rm{DM}}$, tends to be smaller than the theoretical mean $\overline{\rm{DM}}$. Larger values of $M_{\rm{turn}}$ and $\zeta$, which lower $\overline{\rm{DM}}$ due to pushing reionization to lower redshifts, are then favored. With these small FRB sample sizes, our assumption of a Gaussian likelihood in Section \ref{sec:MCMC_setup} limits our ability to extract enough information from $\overline{\rm{DM}}$ to place constraints on the EoR parameters. Taking advantage of the non-Gaussianity of such a measurement might be able to further constrain these parameters, given the same number of high redshift FRBs. 

%Posterior distributions for measurement of $\overline{\rm{DM}}$ for $10$ FRBs in each redshift bin between $ 8 \leq z \leq 10$. The 68$\%$ credibility regions of our measurements are unable to capture the fiducial reionization scenario.

\section{Conclusion}
\label{sec:Conclusion}
The DM of FRBs depend on the free electrons along the line of sight, and so the DM of high redshift FRBs will naturally contain information about the ionization state of the IGM. This makes detection of high redshift FRBs a potentially useful probe to study cosmic reionization. Here we study how the astrophysics and morphology of the EoR affects the mean DM of high redshift FRBs. We use a parametrization, $\beta$, that tracks the density-ionization correlation in the EoR and common astrophysical parameters to bracket the range of physical EoR scenarios. We find that $\overline{\rm{DM}}$ is sensitive to the astrophysics and morphology of reionization and can influence fluctuations in $\overline{\rm{DM}}$ up to $1000$pc cm$^{-2}$. In particular, the ionizing efficiency and mass scale of the ionizing sources cause the greatest fluctuations in $\overline{\rm{DM}}$, which we physically attribute to being caused by the modified timing of reionization. The EoR morphology impacts $\overline{\rm{DM}}$ by changing the density of free electrons within the ionized regions. We find that inside-out reionization scenarios produce the greatest density of free electrons within the ionized bubbles which increases the mean DM of high redshift FRBs with respect to outside-in reionization scenarios. To gauge the viability of such a probe, we perform numerical forecasts to study the types of constraints that can be placed on the astrophysical and correlation parameters using measurements of highly dispersed FRBs. We find that samples of 100 FRBs can rule out uncorrelated reionization at $68\%$CR. Using samples of $10^4$ FRBs in the same redshift range can rule out uncorrelated and outside-in reionization at $95\%$CR. We also find that samples of 100 FRBs between $8  \leq z \leq 10$ can rule out scenarios where the Universe is entirely neutral at $z = 10$ with $68\%$CR. Further, this measurement can also rule out EoR scenarios where the IGM is more than $60\%$ ionized at $z = 8$. Larger sample sizes ($\geq10^4$), of high redshift FRBs, distributed in redshift from $8 \le z \le 10$ according to the theoretically motivated source count distributions, can constrain the duration of reionization (duration between mean ionized fractions 0.25 to 0.75) to $\Delta z = 2.1^{+0.50}_{-0.30}$ and midpoint $z = 7.8^{+0.20}_{-0.20}$ at $68\%$CR. Finally, we find that samples of $\geq10^5$ high redshift FRBs can constrain the duration of reionization (duration between mean ionized fractions 0.25 to 0.75) to $\Delta z = 2.0^{+0.5}_{-0.4}$ and midpoint $z = 7.8^{+0.4}_{-0.2}$ at $95\%$CR.

For future work, we would like to further this proof of concept by using the full distribution of DMs at each $z$ in our forecasts, and by making use of observational constraints as well as the intrinsic constraints outlined in this paper. There are, most obviously, observational constraints that play a role in the feasibility of such parameter fitting with real data.  While high-DM (DM > 4000) events have not yet been observed, \cite{Liam} notes that one can design an experiment that has a higher detection rate of highly dispersed events by trading time resolution for higher frequency resolution. \cite{Zhang+2021} note that FAST and SKA will have the capability of making such detections and most recently \cite{Hashimoto2021} show that observations from SKA phase 2 will indeed reveal our reionization history. It must be noted, however, that the FRB progenitor will ultimately dictate whether there exists an FRB population during the EoR. In addition, a more sophisticated simulation would allow one to explore correlations between $\rm{M}_{\rm{turn}}$, $\zeta$, and the FRB source count distribution since these three parameters ultimately depend on the stellar population. Folding everything into one framework would allow one to study such correlations as well as take clustering of FRBs into account.

Cosmic Dawn did not occur as a single bright event, but rather individual stars, one by one, lit our dark universe. Similarly, we will likely not understand Cosmic Dawn and the Epoch of Reionization from one observation alone, but rather, we will need to make use of many tools which, one by one, will illuminate our understanding of this mysterious time in our universe's history. We propose here that the careful study of highly dispersed FRB observables can serve as such a tool which will, along with many others, help us understand the Epoch of Reionization.

\section*{Acknowledgements}

We would like to extend our deepest thanks to Wenbin Lu for encouraging us to work on this project and for providing us with key insights and guidance. We would also like to thank Adrian Liu and Jordan Mirocha for their helpful comments. There was no explicit funding for this work, though we acknowledge support from the New Frontiers in Research Fund Exploration grant program, a Natural Sciences and Engineering Research Council of Canada (NSERC) Discovery Grant and a Discovery Launch Supplement, the Fonds de recherche du Québec – Nature et technologies (FRQNT), the Sloan Research Fellowship, the William Dawson Scholarship at McGill, as well as the Canadian Institute for Advanced Research (CIFAR) Azrieli Global Scholars program. This research was enabled in part by support provided by Calcul Quebec (\url{www.calculquebec.ca}), WestGrid (\url{www.westgrid.ca}) and Compute Canada (\url{www.computecanada.ca}). 

%%%%%%%%%%%%%%%%%%%%%%%%%%%%%%%%%%%%%%%%%%%%%%%%%%
\section*{Data Availability}
The software code underlying this article will be shared on reasonable request to the corresponding authors.



\bibliographystyle{mnras}
\bibliography{FRB_paper} % if your bibtex file is called example.bib

%\begin{thebibliography}{99}
%\end{thebibliography}

%%%%%%%%%%%%%%%%%%%%%%%%%%%%%%%%%%%%%%%%%%%%%%%%%%

% Don't change these lines
\bsp	% typesetting comment
\label{lastpage}
\end{document}